\documentclass[aps,pre,reprint]{revtex4-2}
\usepackage{amsmath,amssymb}
\usepackage{empheq}
\usepackage{gensymb}

\begin{document}

\title{Asymmetric Assembly of Lennard-Jones Janus Dimers}
\author{Sina Safaei,$^{1, 2}$ Caleb Todd,$^2$ Jack Yarndley,$^2$ Shaun Hendy,$^{1, 2, 3}$ and Geoff R. Willmott$^{1, 2, 4}$}
\affiliation{$^1$The MacDiarmid Institute for Advanced Materials and Nanotechnology, New Zealand\\
$^2$Department of Physics, University of Auckland, New Zealand\\
$^3$Te P\={u}naha Matatini, Department of Physics, University of Auckland, New Zealand\\
$^4$School of Chemical Sciences, University of Auckland, New Zealand}

\begin{abstract}
\noindent Self-assembly of Janus (or `patchy') particles is dependent on the precise interaction between neighbouring particles. Here, the orientations of two amphiphilic Janus spheres within a dimer in an explicit fluid are studied with high geometric resolution. Molecular dynamics simulations and first-principles energy calculations are used with hard- and soft-sphere Lennard-Jones potentials, and temperature and hydrophobicity are varied. The most probable centre-centre-pole angles are in the range $40\degree$ to $55\degree$, with pole-to-pole alignment not observed due to orientational entropy. Angles near $90\degree$ are energetically unfavoured due to solvent exclusion, and the relative azimuthal angle between the spheres is affected by solvent ordering. Relatively large polar angles become more favoured as the hydrophobic surface area (i.e. Janus balance) is increased.
\end{abstract}

\maketitle

\section{Introduction}

Janus particles (JPs) are micro- or nanoparticles that have at least two sides with distinct physical or chemical properties \cite{bergstrom2011thermodynamics,yang2012janus,poggi2017janus,popescu2020chemically,perro2005design}. JPs can have asymmetry in (for example) optical, electrical, or magnetic properties; they can be synthesised with different shapes (e.g. spheroids, rods, platelets); and properties can be altered over particular areas, resulting in `patchy' particles. Amphiphilic microspheres, which have different wettability on each hemisphere, are a commonly-studied type of JP \cite{walther2008janus,walther2013janus,chen2011supracolloidal,roh2005biphasic,jiang2010janus,preisler2014equilibrium}. 

Directional interactions between JPs can lead to self-assembly of complex structures \cite{poggi2017janus,popescu2020chemically,safaie2020janus}, and there has been extensive interest in related design rules for Janus and patchy particles \cite{zhang2017janus,zhang2015toward,hagan2011mechanisms,grzybowski2017dynamic}. In 3D, experiments have produced aggregates ranging from small clusters up to non-equilibrium helices with variable chirality \cite{chen2011supracolloidal}. Simulations have predicted phases including small clusters \cite{rosenthal2012self}, micellar or wire-like structures \cite{munao2013cluster,rosenthal2012self,li2012model,hu2019enthalpy,moghani2013self}, and sheets \cite{preisler2014equilibrium,munao2013cluster}. In 2D, close-packed particles \cite{baran2020self,kretzschmar2011surface} can produce tiled patterns \cite{iwashita2014orientational,shin2014theory,chen2012janus}, while less dense arrangements include clusters \cite{chen2012janus}, chains \cite{iwashita2013stable} and open lattices \cite{mao2013entropy}. Phases depend on parameters such as solvent properties, temperature, and patch geometry \cite{miller2009hierarchical,moghani2013self}. 

An analogy can be made between particle self-assembly and formation of chemical bonds, albeit with different rules governing the geometry of particle clusters \cite{sciortino2019entropy, zhang2015toward,wang2012colloids} that are yet to be fully understood. For example, the number and directions of `bonds' can be controlled via the positions and sizes \cite{sciortino2019entropy,zhang2015toward,chen2012janus,mao2013entropy,zhang2017janus,grzybowski2017dynamic} of patches. Orientational entropy determines the flexibility (or `floppiness') of bonds \cite{iwashita2014orientational,mao2013entropy,sciortino2019entropy,zhang2015toward,chen2011supracolloidal}, and can determine the stability of a phase \cite{wang2012colloids,chen2011directed,mao2013entropy,sciortino2009phase,smallenburg2013liquids,sciortino2019entropy}.

The precise nature of interactions between particles ultimately determines the possible aggregates. Simulations are an important predictive tool in this field \cite{zhang2015toward}, and a variety of anisotropic directionally-dependent point potentials have been used to study JP assembly \cite{baran2020self}. These potentials typically consist of an isotropic repulsion such as an infinitely large `hard-sphere' potential barrier \cite{hong2008clusters,kern2003fluid,preisler2014equilibrium,munao2013cluster}, as well as an anisotropic interaction representing the JP asymmetry. Kern and Frenkel's approach \cite{kern2003fluid}, which has been widely used \cite{mao2013entropy,iwashita2013stable,iwashita2014orientational,shin2014theory,preisler2014equilibrium,munao2013cluster,baran2020self}, applies a square-well potential when attractive surfaces are in contact on the line joining the centres of two particles, and zero otherwise. This is suitable for modelling short-range forces, and produces an energetic step function with respect to JP orientation at the boundary of a hemisphere (or patch). As an alternative, the energy can be minimised when attractive hemispheres are in pole-to-pole contact, and increase according to vector dot products as the JPs are rotated \cite{li2012model,rosenthal2012self,hong2008clusters,baran2020self}. These potentials take reasonable simple forms for the purpose of many-body calculations. There are limited examples of more detailed geometric calculations of the energy between two JPs, in which the interaction has been based on electrostatics (i.e. DLVO theory) \cite{hong2006clusters,hieronimus2016model} and critical Casimir forces \cite{squarcini2020critical,labbe2016critical}.

In this paper, we study the 3D orientation dynamics of amphiphilic spherical JPs in self-assembled dimers. We use a Lennard-Jones potential and two geometrically rigorous calculation methods; none of these features have appeared in previous studies of the interaction between two JPs. Results obtained using molecular dynamics (MD) simulations of many-atom JPs are presented alongside calculations made using a numerical integration (NI) method we have developed. In the latter method, JP hemispheres are modelled as continuous surfaces, and potentials are integrated to calculate the first-principles interaction energy for any particular configuration. An analytic equivalent of our NI method has been used to model other Lennard-Jones nanostructures \cite{baowan2016mathematical} including uniform spheres \cite{baowan2017modelling}, but to our knowledge this has not been applied to Lennard-Jones JPs. 

Put together, our results produce a detailed and nuanced description of particle interactions for short-ranged potentials, and specifically the Lennard-Jones potential which approximates van der Waals forces. The MD and NI approaches both directly include solvent, have non-zero separation of JP and solvent surfaces, and can account for orientational entropy. In addition, the MD simulations allow us to study solvent ordering, changing surface separation lengths, and other dynamics. Results from MD and NI are broadly consistent with each other at different temperatures and hydrophobicities, enabling discussion of the geometric and entropic contributions to the preferred orientations of JPs in a dimer. Aside from improving fundamental understanding in this way, the results can inform the development of design rules in self-assembled structures using `floppy' bonds. Dimers of JPs are particularly important because they are the first kinetic step in any self-assembly \cite{chen2011supracolloidal}, they are possible stand-alone modular building blocks, and they appear as a motif in 2D tiling \cite{shin2014theory,iwashita2014orientational,chen2012janus}.

\begin{figure}[b]
    \includegraphics[width=0.6\linewidth]{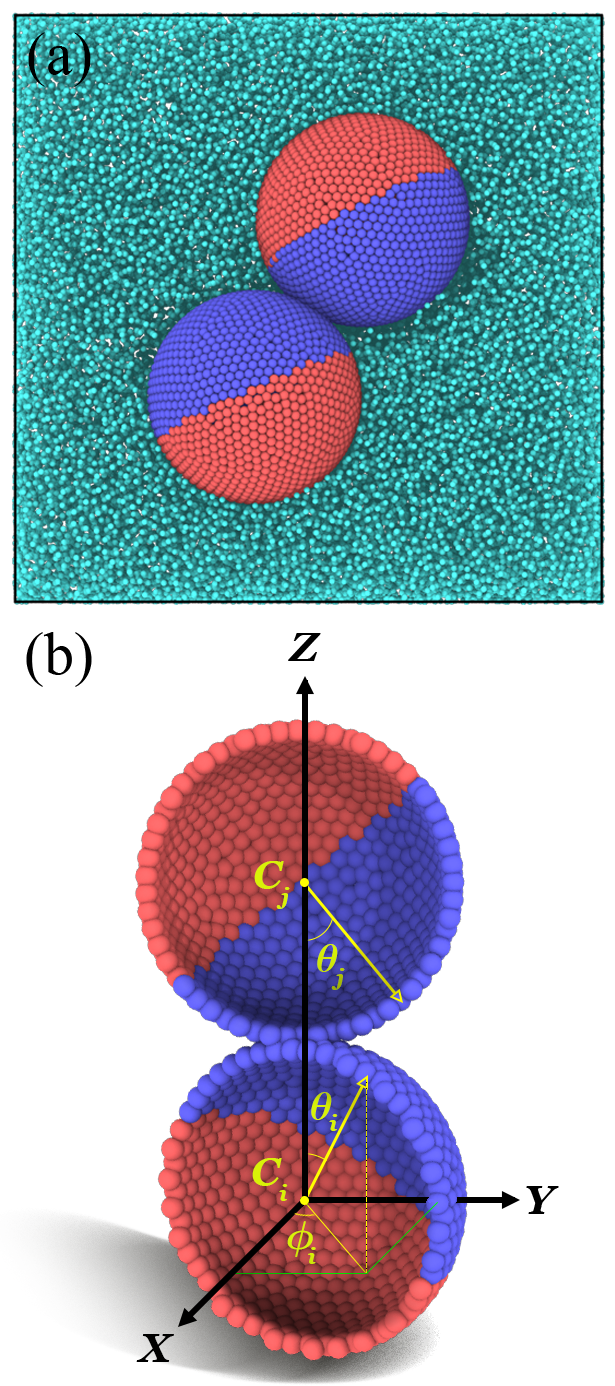}
    \caption{(a) Simulation setup for a Janus dimer ($R=10~\sigma$) inside a fluid box with side lengths of $50~\sigma$. (b) Cross-section of Janus spheres showing the polar angles $\theta_i$ and $\theta_j$ and the azimuthal angle $\phi_i$. The angle $\phi_j=90\degree$ is omitted for clarity. $C_i$ and $C_j$ are the centres of the spheres, and the vectors connect the centres to hydrophobic poles.}
    \label{Fig_Box_Angles}
\end{figure}

\section{Molecular dynamics}

MD simulations were carried out using the large-scale atomic molecular massively parallel simulator (LAMMPS), and the open visualisation tool (OVITO) was employed for visualisation and analysis of the results \cite{plimpton1995fast,stukowski2009visualization}. All simulations were performed in Lennard-Jones reduced units, where $\sigma$ is the unit of distance, $\epsilon$ the unit of energy, $m$ the unit of mass and $\tau$ the unit of time ($\sqrt{m\sigma/\epsilon}$), without loss of generality. The equations of motion were integrated using the standard velocity-Verlet algorithm with a timestep of $\Delta t=0.01~\tau$. All simulations were carried out in a canonical ensemble (NVT) in which the temperature was set at 1.1 $\epsilon/k_B$ and controlled by a Langevin thermostat with a damping parameter of $10~\tau$.

MD studies of JPs can typically be categorised into those which model Janus particles as single `atoms' using point potentials \cite{rosenthal2012self,hu2019enthalpy,vissers2013predicting,baran2020self} or those which represent spherical JPs as collections of atoms distributed over the surface of hollow spheres \cite{li2019computer,xu2015self,kobayashi2020structure,safaei2020stability,safaei2019molecular}. A many-atom JP can be used to rigorously study hydrodynamics such as the effect of slip on particle rotation and translation \cite{willmott2008dynamics,willmott2009slip,safaei2019molecular,kharazmi2015diffusion}. Therefore, Janus spheres were modelled as in our previous studies \cite{safaei2019molecular,safaei2020stability}, where the atoms were distributed over the sphere surfaces with a density of $1.3~\sigma^{-2}$. The positions of atoms were randomised for each simulation, avoiding artefacts associated with regular structures (discussed in Sec. \ref{sec:4c}). Anisotropy of JPs emerges by defining different potentials for atoms on different parts of the surface.
Hence, the spheres were divided geometrically into two hemispheres referred to as $\mathbf{1}$ and $\mathbf{2}$, and they interacted with atoms of an explicit solvent referred to as $\mathbf{0}$. To design the spheres as similarly as possible to experimentally fabricated JPs, the boundary between the hemispheres is not sharp. This eliminates any related effects \cite{safaei2019molecular} and means that the number of atoms on different hemispheres may not be equal.

In this study, a periodic simulation box with side lengths of $a = 50~\sigma$ was designed in which two Janus spheres (radius $R = 10~\sigma$) are in a solvent with density of $\rho = 0.75~\sigma^{-3}$ (see Fig. \ref{Fig_Box_Angles}(a)). Thus, the minimum possible distance between a Janus dimer and its images is more than three times the cut-off radius ($\approx 10 \sigma$). The temperature and density of the solvent atoms were chosen in a way to ensure the solvent is a fluid phase \cite{wang2020lennard}. To start each simulation, the fluid was relaxed for a million timesteps at the desired temperature (default value $1.1~\epsilon/k_B$) while the spheres were fixed. Then, the simulation continued for another million timesteps in which the spheres were free to move and came to contact on their hydrophobic sides after a short period. Afterwards, the simulation continued for $100$ million timesteps for data collection. $25$ independent simulations were performed for each set of parameters, each with newly randomised atom positions on the surfaces of the spheres to ensure the result is not affected by the sphere design. 

To study the orientation of individual JPs in a dimer, two spherical coordinate systems $(r_{i,j},~\theta_{i,j},~\phi_{i,j})$ were employed (Fig.~\ref{Fig_Box_Angles}(b)), where the origins are at the centres of each of the spheres $C_{i,j}$, and the line connecting the centres ($\overrightarrow{C_iC_j}$) indicates the direction of the $z$-axis. The orientation of each sphere is defined by a vector connecting its centre to its hydrophobic pole. The axes are defined so that $\phi_j = 90\degree$ and the relative orientation of the two Janus spheres can be fully specified using $\theta_i,~\theta_j$, and $\Delta\phi$ (where $\Delta\phi=\phi_j-\phi_i$). The Supplemental Material \cite{SM} includes illustrations of the orientations of Janus spheres at different polar and azimuthal angles to aid visualisation of variations of these parameters.

The potential between atoms separated by distance $r$ was a 12-6 Lennard-Jones potential:

\begin{equation}\label{Eq:LJ} 
U_{ij}\left(r\right)=4 \epsilon \left[ \left(\frac{\sigma}{r}\right)^{12} - \left(\frac{\sigma}{r}\right)^{6} \right], \\
\end{equation}

\noindent with the interaction strengths between different types of atoms defined as:

\begin{empheq}[left = \empheqlbrace]{align}
    &\epsilon_{(1-0)} = A_o\epsilon_{(0-0)} \nonumber \\
    &\epsilon_{(2-0)} = A_i \epsilon_{(0-0)} \nonumber \\
    &0.05 \leq A_o \leq A_i \leq 1.00 \nonumber.
\end{empheq}

\noindent Here $A$ (the `interaction strength') is a parameter allowing the potential to be adjusted for the hydroph\textbf{\underline{o}}bic~($A_o$) and hydroph\textbf{\underline{i}}lic~($A_i$) hemispheres in a Janus sphere, which are depicted in all figures as blue and red, respectively. As the interaction strength increases, the hydrophobicity of the surface decreases. The interactions $\epsilon_{(1-1)}$, $\epsilon_{(2-2)}$, and $\epsilon_{(1-2)}$ were calculated using the Berthelot mixing rule. Parameters $m$ (mass) and $\sigma$ were the same for all atom types.

\section{Numerical integration}

\begin{figure}[h]
    \includegraphics[width=0.4\linewidth]{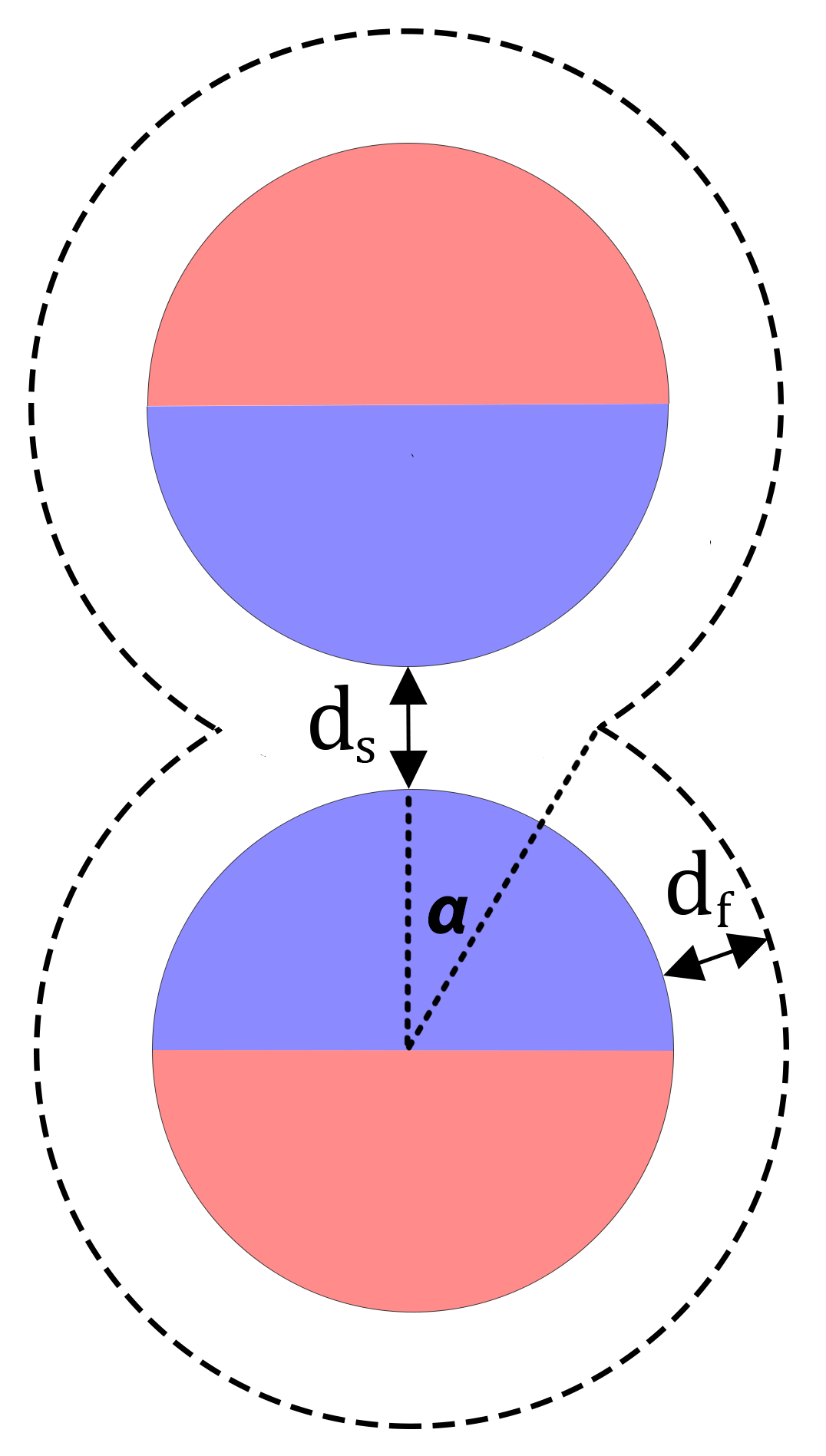}
    \caption{A Janus dimer, showing the parameters $d_s$ and $d_f$ used in NI method. $\alpha$ is the angle at which fluid exclusion zones overlap.}
    \label{Fig_2D_Schem}
\end{figure}

In the numerical integration (NI) approach, we calculated the interaction between two Janus spheres over a range of configurations. Each sphere consisted of two continuous hemispherical surfaces separated by a straight 
boundary along the equator (Fig.~\ref{Fig_2D_Schem}). This contrasts with the discrete atoms and imperfect boundaries between hemispheres used in MD. The centres of the two spheres were positioned at $(0,0,0)$ and $(0,0,2R+d_s)$ in Cartesian coordinates, where $d_s$ is the distance between the spheres' surfaces. A fixed distance between the solid surfaces and the solvent ($d_f$) was also defined. In MD, $d_s$ and $d_f$ continuously change in response to the configuration, and $d_f$ is (on average) different for the hydrophilic and hydrophobic sides. Default values of $d_s = 0.51\sigma$ and $d_f = 0.8\sigma$ were chosen by extracting averages from MD, but calculations were also made using several different values to show that these parameters have a minimal effect on orientation probability distributions (see Supplemental Material \cite{SM}). Otherwise, NI calculation parameters (e.g. surface and fluid densities) matched the values used in MD simulations. 

The energy at any given configuration was calculated by integrating the Lennard-Jones interaction over the different geometries, resulting in a total of 8 integrals (4 surface-surface and 4 surface-solvent). The form of the integrals between the different geometries is either:

\begin{equation}
  U_1 = \iint_{S_1} \iint_{S_2} U_{LJ}(r) \; dS_2 dS_1,
  \label{int1}
\end{equation}

\noindent where $S_1$ and $S_2$ are the surface areas of different hemispheres, or:

\begin{equation}
  U_2 = \iint_{S} \iiint_{V} U_{LJ}(r) \; dV dS,
  \label{int2}
\end{equation}

\noindent where $S$ is the area of a hemisphere and $V$ denotes the solvent's volume. 

The integrals were calculated using spherical coordinates local to each sphere for ease of parameterisation. To find the distance between points in two distinct coordinate systems, we convert the local spherical coordinates of each point into a global Cartesian system by applying standard rotation and translation operations. We then compute the Euclidean distance between the points. In the evaluation of Eq. \ref{int2}, the fluid domain excludes a lens shaped volume formed by the overlapping regions in which solvent is excluded adjacent to the sphere surfaces (Fig. \ref{Fig_2D_Schem}). The volume integral is first computed as though there were no overlap. 
Then, the lens-shaped overlapping zone was independently parameterised and its interaction was subtracted from the total energy.  

Numerical integration was performed using the Julia programming language \cite{Julia-2017} with the HCubature.jl package. For a particular configuration defined by $\theta_i$, $\theta_j$, and $\phi$, the energy is $E(\theta_i, \theta_j, \phi)$ and the corresponding configuration probability density can be determined using

\begin{equation}
    P = \frac{1}{Z}\iiint g\exp\left[-\frac{E}{k_BT}\right]d\theta_j'd\theta_i'd\phi'
\end{equation}

\noindent where $Z$ is the partition function, $k_B$ is Boltzmann's constant, $T$ is temperature and $g = g(\theta_i, \theta_j, \phi)$ is the density of states for the configuration. Here $\phi$ is equivalent to the relative azimuthal angle ($\Delta\phi$) used elsewhere in this paper. For a sphere at a given orientation, rotation about the polar vector does not change the state, and hence the density of states for sphere $i$ is proportional to $2 \pi R \sin(\theta_i)$, i.e. the circumference of the circle traced by the contact point of the two spheres during the rotation. Thus, the number of ways a configuration of $\theta_i, \theta_j$ can be obtained is proportional to $\sin(\theta_i)\sin(\theta_j)$ and the salient features of the configuration probability density are contained in

\begin{equation}
    p \propto \sin(\theta_i)\sin(\theta_j)\exp\left[-\frac{E}{k_BT}\right].
    \label{ProbCalc}
\end{equation}

In addition to the hard-sphere (12-6) potential in Eq.~\ref{Eq:LJ}, we carried out calculations using a soft-sphere (9-6) Lennard-Jones potential,

\begin{equation}
U_{ij}\left(r\right)=6.75 \epsilon \left[ \left(\frac{\sigma}{r}\right)^{9} - \left(\frac{\sigma}{r}\right)^{6} \right]. \\
\end{equation}

\noindent This was of interest due to frequent use of soft directional-dependent potentials in previous studies of Janus spheres \cite{li2012model,li2013simulation,li2014soft,li2016versatile,li2016supracolloidal,li2018general,zou2016supracolloidal,zou2019coupling}. 

\section{Results and discussion}

\subsection{Polar angle}

\begin{figure}[h]
    \includegraphics[width=0.9\linewidth]{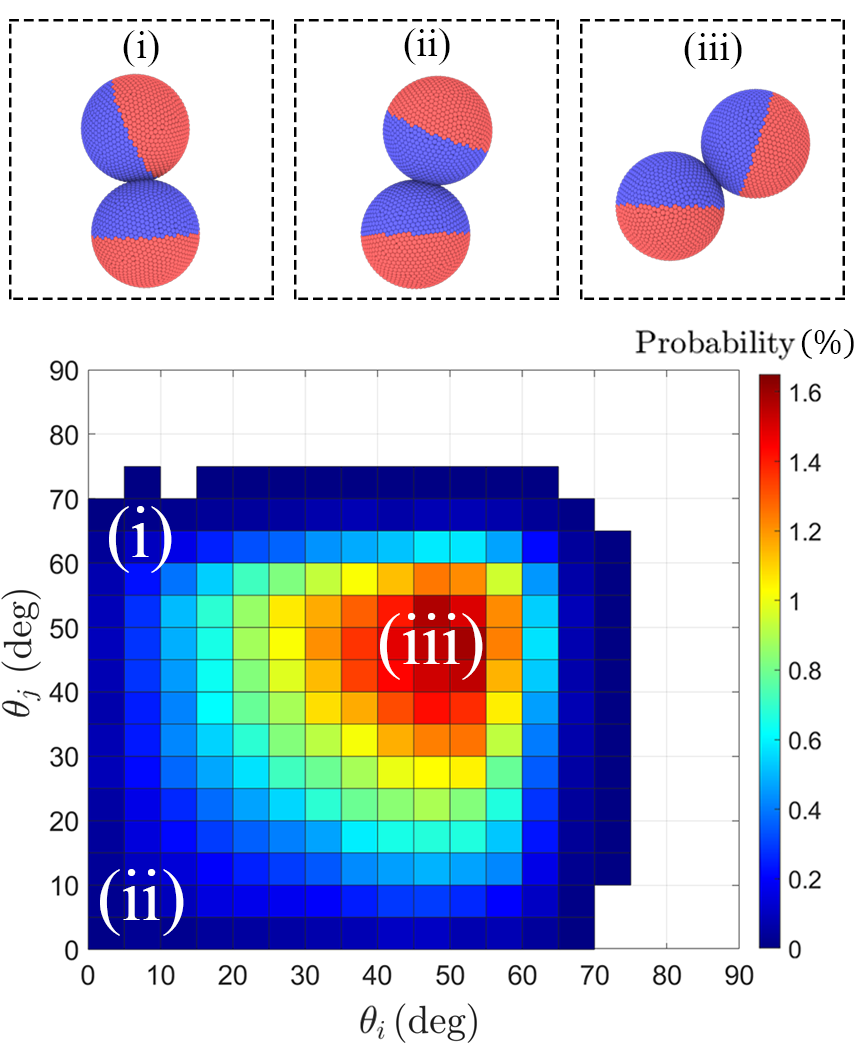}
    \caption{The occurrence probability of $\theta_i$ and $\theta_j$ at $T=1.1~\epsilon/k_B$, $A_o=0.05~\epsilon$, and $R=10~\sigma$. Insets illustrate sphere orientations at specified points on the plot.}
    \label{Fig_Polar_Hist}
\end{figure}

Figure~\ref{Fig_Polar_Hist} is an example of an occurrence probability distribution as a function of the polar angles $\theta_i$ and $\theta_j$, simulated using MD. 
This plot shows the most important general features of the JP dimer orientation probability distributions in this study. The observed angles are most commonly in the range from $40\degree$ to $55\degree$, and orientations with angles higher than $75\degree$ are not observed. Orientations with angles less than $30\degree$ are not favoured due to the lower available surface area at those angles - there are relatively few ways such a configuration can be obtained (see Eq.~\ref{ProbCalc}). This is the effect of orientational entropy. The sudden decrease above $60\degree$ is related to screening of solvent atoms from the hydrophobic hemispheres, which will be discussed below. The following sections explore variations in data comparable to Fig.~\ref{Fig_Polar_Hist} due to the modelling approach and type of potential used, particle size, temperature and hydrophobicity, and the geometric design of the Janus spheres.

\subsubsection{Models and potentials}

\begin{figure}[h]
    \includegraphics[width=0.9\linewidth]{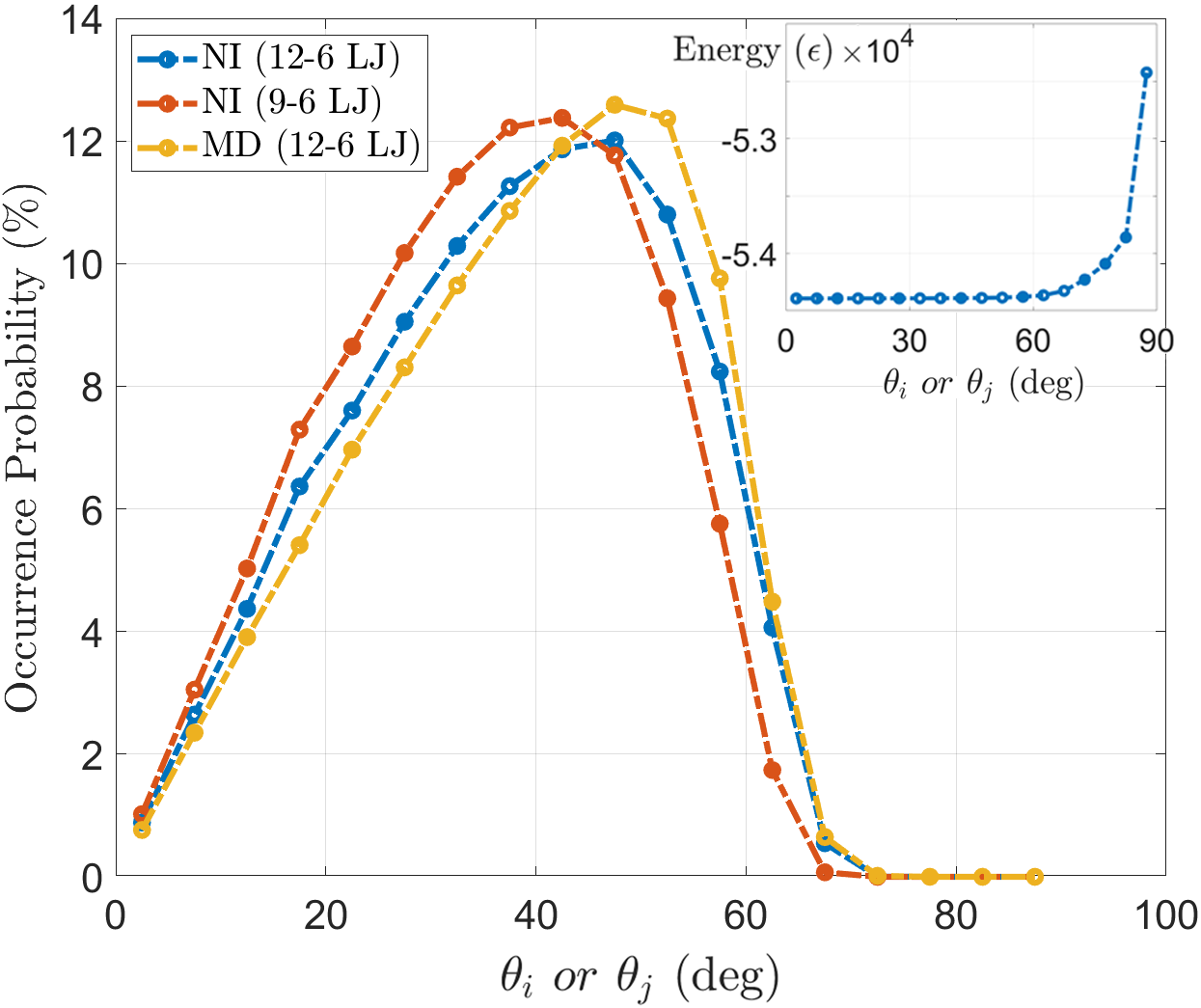}
    \caption{The occurrence probability of orientations with different $\theta$ angles for Janus spheres in a dimer ($T=1.1~\epsilon/k_B$, $A_o=0.05~\epsilon$, $R=10~\sigma$). The inset shows the energy landscape of the sphere configurations as a function of polar angle. Lines are drawn to guide the eye, and error bars are smaller than the markers.}
    \label{Fig_Potential_Eff}
\end{figure}

Fig. \ref{Fig_Potential_Eff} compares the occurrence probability of polar angles calculated using MD and NI methods. The MD simulations show a good agreement with NI calculations with a matching hard-sphere potential, where the maximum difference between the plots is $1.5\%$ at $55\degree$. In particular, the maximum probability occurs at very similar angles ($\approx5\degree$ difference). The likely reasons for discrepancies between these plots are the known differences between the MD and NI models, i.e. the fixed values of $d_s$ and $d_f$ and the sharp boundary between hemispheres in the NI model. For the soft-sphere potential, the probability distribution shifts slightly toward smaller angles. 

To further understand these probability distributions, consider the calculated energy for the interaction between the two spheres (inset to Fig.~\ref{Fig_Potential_Eff}) which is approximately constant for all configurations with $\theta_{i,j} < 60\degree$. In this region, the reason for variation in occurrence probability is the lower orientational entropy (due to smaller surface area) near the poles. This is characteristic of potentials which have little or no dependence on orientation, representing short-range interactions \cite{shin2014theory}. In contrast, approaches which use geometrically rigorous long-range interactions (electrostatic or DLVO) produce continuously varying potential landscapes \cite{hong2006clusters,hieronimus2016model}.

At $\theta_{i,j}>60\degree$, the interaction energy gradually increases, producing a sharp decline in the probability distribution for $60\degree<\theta_{i,j}< 90\degree$. This decline is not present when using a Kern-Frenkel type potential. In those cases, the maximum probability is found adjacent to the step function (at $\theta_{i,j} \approx 90\degree$ for hemispheres), where the energy abruptly increases and the orientational entropy is greatest. Labb\'{e}-Laurent and Dietrich \cite{labbe2016critical} derived a geometrically rigorous potential for a Janus dimer interacting via Casimir forces, which produced a similar combination of constant and varying potentials as in Fig.~\ref{Fig_Potential_Eff} inset, albeit without solvent. 

The main reason for the increase in energy over the range $60\degree<\theta_{i,j}< 90\degree$ is that solvent is excluded from the region near the point of closest approach between the spheres. It is energetically favourable for the hydrophobic surfaces to be shielded from the solvent near this contact point. The volume in which there is no fluid (an `exclusion zone') is determined by $d_f$ (Fig. \ref{Fig_2D_Schem}). The polar angle at which a sphere's hydrophilic side enters this exclusion zone (and thus the hydrophobic side is not optimally hidden from the fluid) is:

\begin{empheq}[left = \empheqlbrace]{align}
    &\theta_{i,j} = 90\degree - \alpha \nonumber \\
    &\alpha = \arccos(\frac{R+d_s/2}{R+d_f}), \nonumber
\end{empheq}

\noindent or $\theta_{i,j} = 72\degree$ for our parameters, which matches well with Figs.~\ref{Fig_Polar_Hist} and \ref{Fig_Potential_Eff} (see Supplemental Material \cite{SM} for a further comparison). The precise shape of these distributions additionally depends on the form and range of the interaction potential.

The shift of the probability distribution to lower polar angles when using a soft-sphere (9-6) potential can be explained by the increased importance of long range interactions. For example, the minimum of the soft potential occurs at a larger particle separation than for the 12-6 potential. Increasing the range of the potential decreases the polar angle required for the hydrophilic side to be affected by the exclusion zone, thereby producing a decline in probability at lower polar angles.

\begin{figure}[h!]
    \includegraphics[width=1\linewidth]{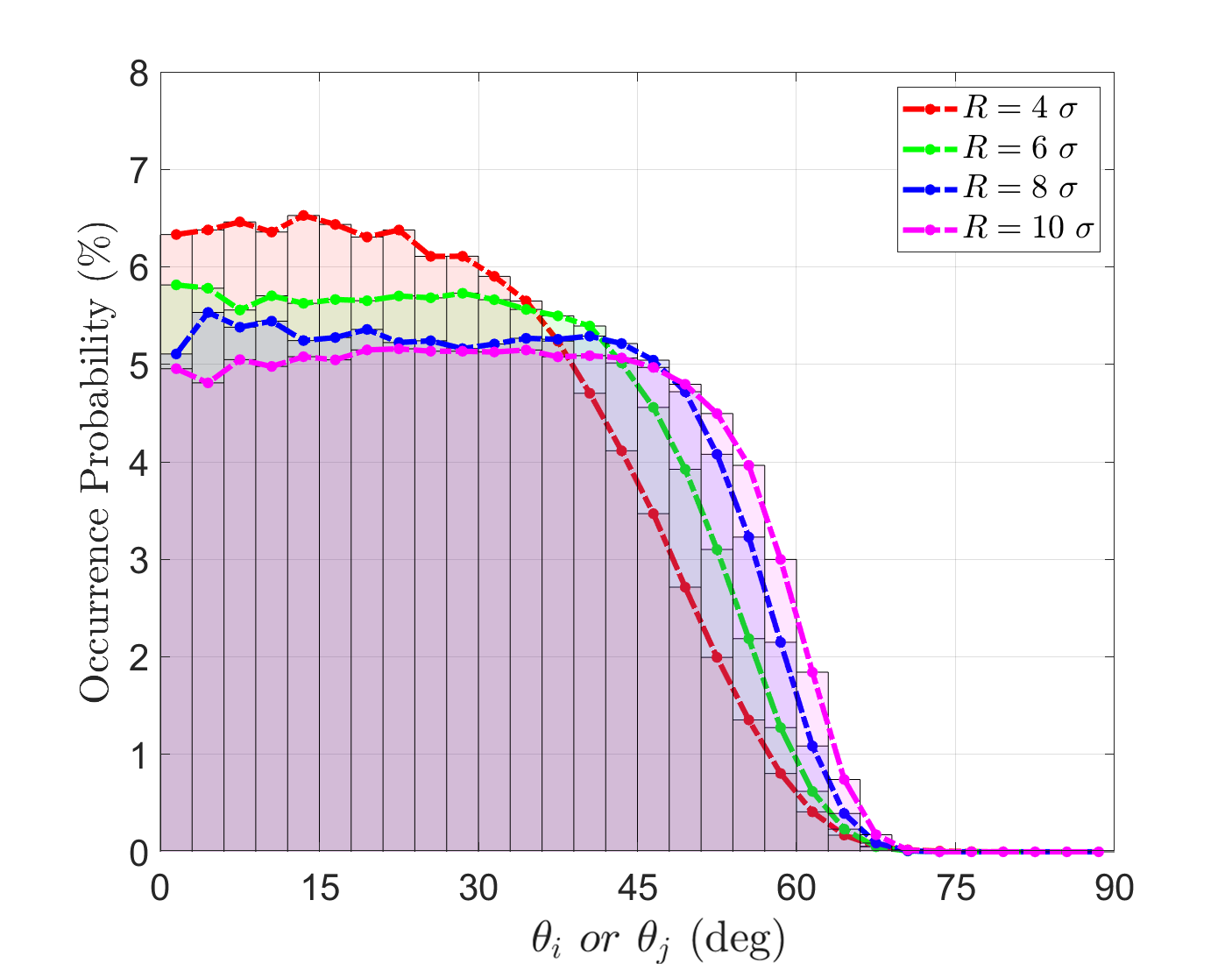}
    \caption{The normalised occurrence probability of $\theta$ for Janus spheres with radii of $R = 4, 6, 8, 10~\sigma$ in a dimer ($T=1.1~\epsilon/k_B$, $A_o=0.05~\epsilon$). The lines are drawn to guide the eye. Error bars are smaller than the markers.}
    \label{Fig_Radius_Eff}
\end{figure}

\subsubsection{Size effect}

To study the effect of sphere size in MD, we designed Janus spheres with radii $R =$~4, 6, 8, and $10\sigma$. Figure~\ref{Fig_Radius_Eff} compares the occurrence probability of $\theta_i$ or $\theta_j$ for spheres with different radii, where the histogram is normalised by the surface area on the spheres corresponding to each histogram bin. This result suggests that the configurations with smaller polar angles are more likely to be observed for smaller spheres. However, the occurrence probability of angles smaller than $\theta\approx45\degree$ is approximately the same for $R=8$ and $10\sigma$. A possible reason for variations with respect to $R$ is the potential cut-off distance, which is comparable to the sphere size. Unfavourable hydrophilic-hydrophobic and hydrophilic-hydrophilic interactions between spheres occur within the cut-off distance. These interactions cause the spheres to avoid orientations with high polar angles. Consequently, smaller spheres spend more time in low-angle orientations than larger spheres. Nonetheless, the occurrence probability drops to zero at angles higher than $\theta\approx70\degree$ regardless of the sphere size due to solvent exclusion. 

\subsubsection{Temperature effect}
\label{Sec:Temp}

\begin{figure}[h]
    \includegraphics[width=1\linewidth]{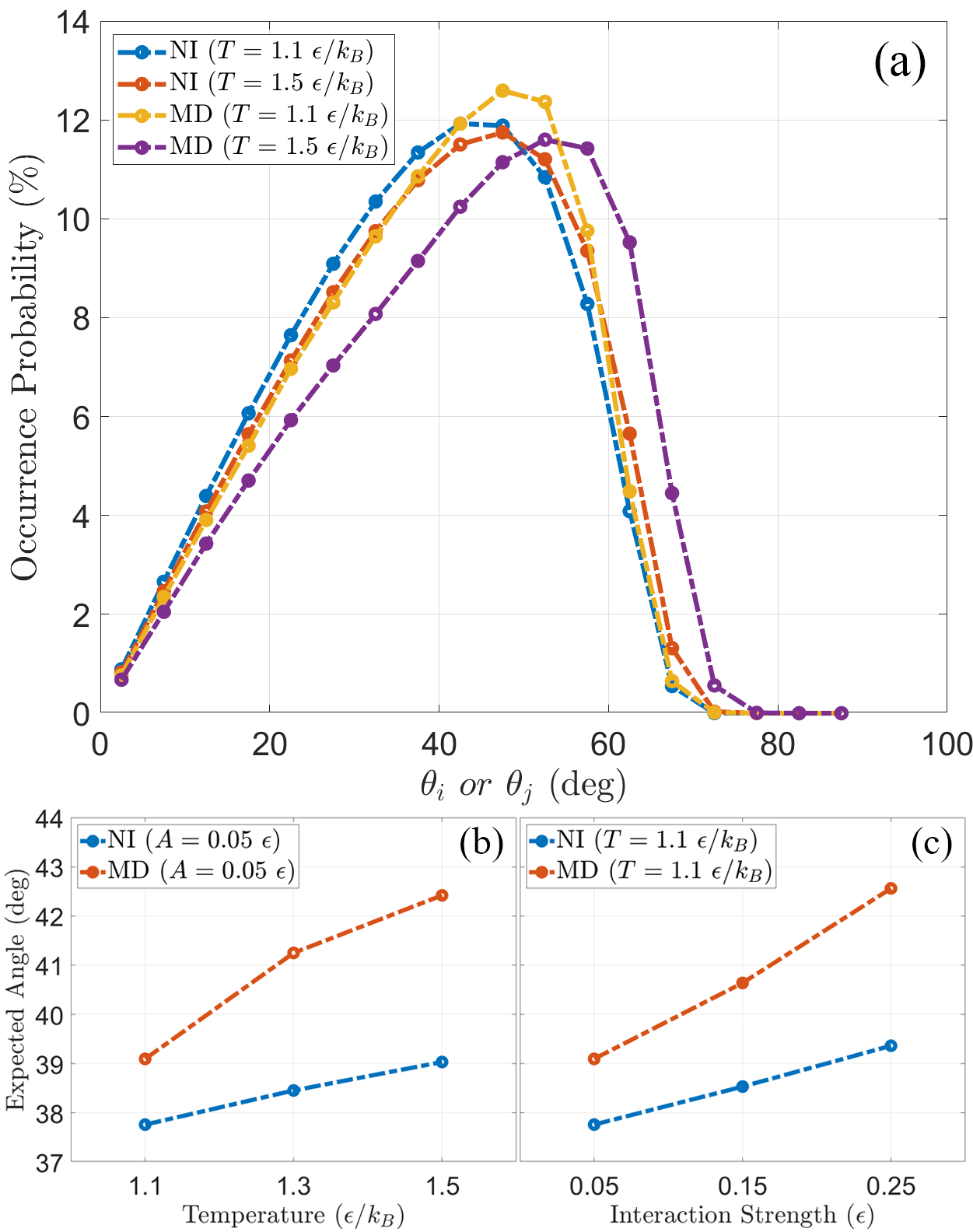}
    \caption{(a) Occurrence probability as a function of polar angle for Janus spheres in a dimer at different temperatures ($A_o=0.05~\epsilon$, $R=10~\sigma$). Lines are drawn to guide the eye. (b) and (c), expected value of the polar angle as a function of temperature ($A_o=0.05~\epsilon$, $R=10~\sigma$) and hydrophobicity ($T=1.1~\epsilon/k_B$, $R=10~\sigma$), respectively. Error bars are smaller than the markers.}
    \label{Fig_Temp_Eff}
\end{figure}

Temperature was varied between $1.1~\epsilon/k_B$ and $1.5~\epsilon/k_B$ to study the effect on JP orientation. Fig.~\ref{Fig_Temp_Eff}(a) shows polar angle probability distributions for two different temperatures calculated using both MD (via repeated simulations) and NI (using Eq.~\ref{ProbCalc}). At higher temperatures, the spheres spend more time in high energy orientations ($\theta_{i,j}>60\degree$) even up to unfavourable angles between $75\degree$ and $80\degree$. Figure~\ref{Fig_Temp_Eff}(b) indicates the expected value of $\theta_{i,j}$ at different temperatures. The results and trends from the MD and NI methods match closely at $1.1~\epsilon/k_B$, where the maximum difference between the plots is $1.5\%$ at $55\degree$. However, the discrepancy increases with temperature to a maximum of $3.9\%$ at $65\degree$. The difference between NI and MD results is possibly caused by fluid layering in MD, as discussed below.

\begin{figure*}
    \includegraphics[width=1\linewidth]{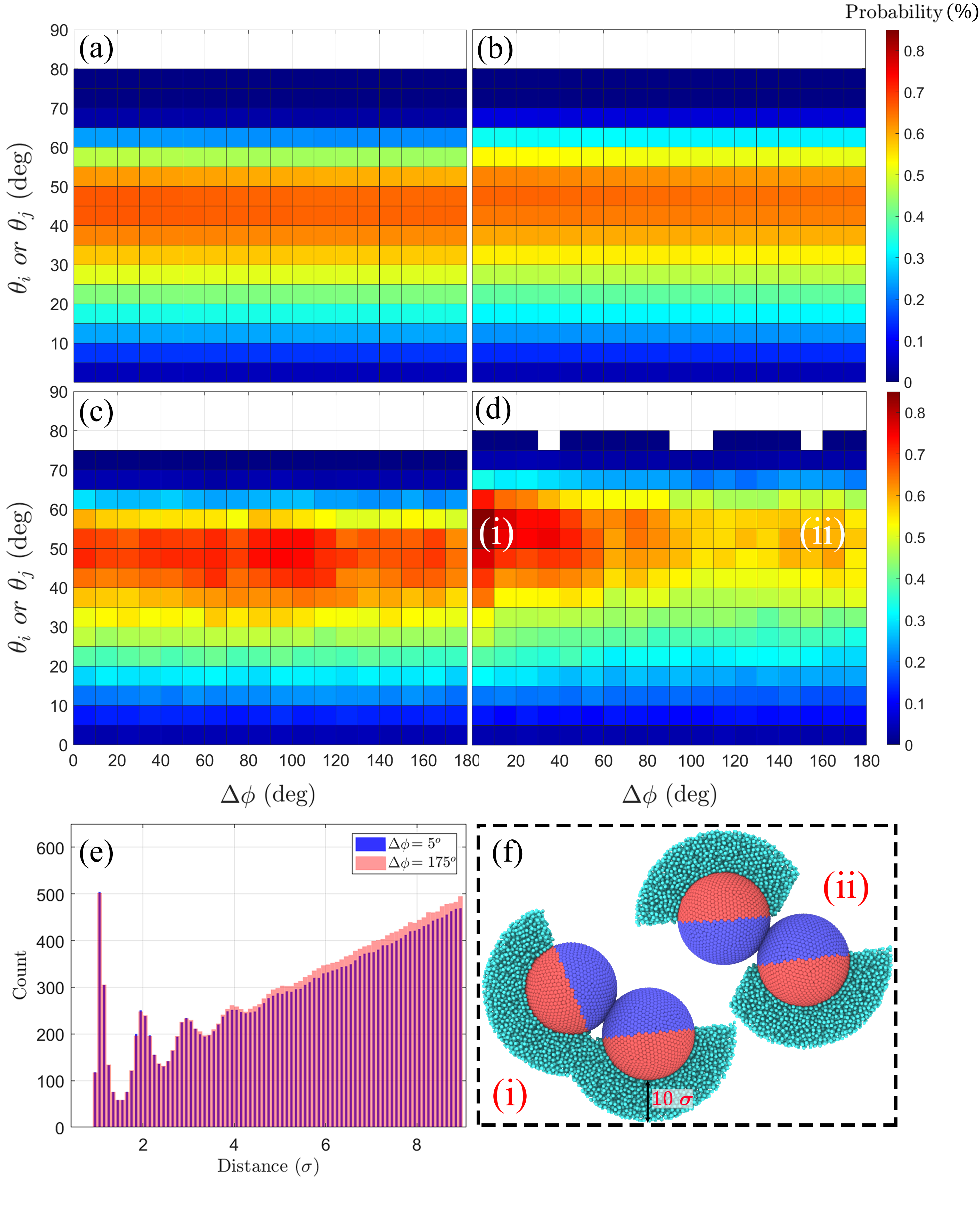}
    \caption{The occurrence probability for dimer orientations ($A_o=0.05~\epsilon$, $R=10~\sigma$) in NI calculations at (a) $T = 1.1~\epsilon/k_B$, and (b) $T = 1.5~\epsilon/k_B$; and in MD simulations at (c) $T = 1.1~\epsilon/k_B$, and (d) $T = 1.5~\epsilon/k_B$. (e) The spatial distribution of solvent atoms near the hydrophilic sides of JPs at $\Delta\phi=5\degree$ and $\Delta\phi=175\degree$. (f) The orientations from (d) at the corresponding labels, illustrating regions of layered solvent atoms used in (e). (ii) does not cover a full semi-circle to maintain equivalence with (i), which has an overlapping region.}
    \label{Fig_Phi_Study}
\end{figure*}

\subsubsection{Hydrophobicity effect}

Previous studies have shown that changing the attraction strength between Janus particles can alter the size and shape of self-assembled nanostructures \cite{miller2009hierarchical,li2019computer,li2012model}. These alterations arise because the favoured orientations of individual Janus spheres in aggregates slightly change. In order to study the effect of hydrophobicity on orientation, the hydrophobic interaction strength was varied from $A_o=0.05\epsilon$ to $0.25\epsilon$ at the default temperature ($1.1~\epsilon/k_B$). The resulting changes in the overall probability distribution (see Supplemental Material \cite{SM}) are very similar to the variations observed for changes in temperature (Fig.~\ref{Fig_Temp_Eff}(a)). Similar trends are also observed in the expected value of $\theta_{i,j}$ (Fig.~\ref{Fig_Temp_Eff}(c)) albeit for a different reason. As the hydrophobicity decreases, the energy of the fluid-hydrophobic interactions becomes more negative which allows the hydrophobic sides of the spheres to be more exposed to the solvent atoms. Nonetheless, the spheres avoid forming orientations with polar angles higher than $70\degree$.

\begin{figure}[h!]
    \includegraphics[width=1\linewidth]{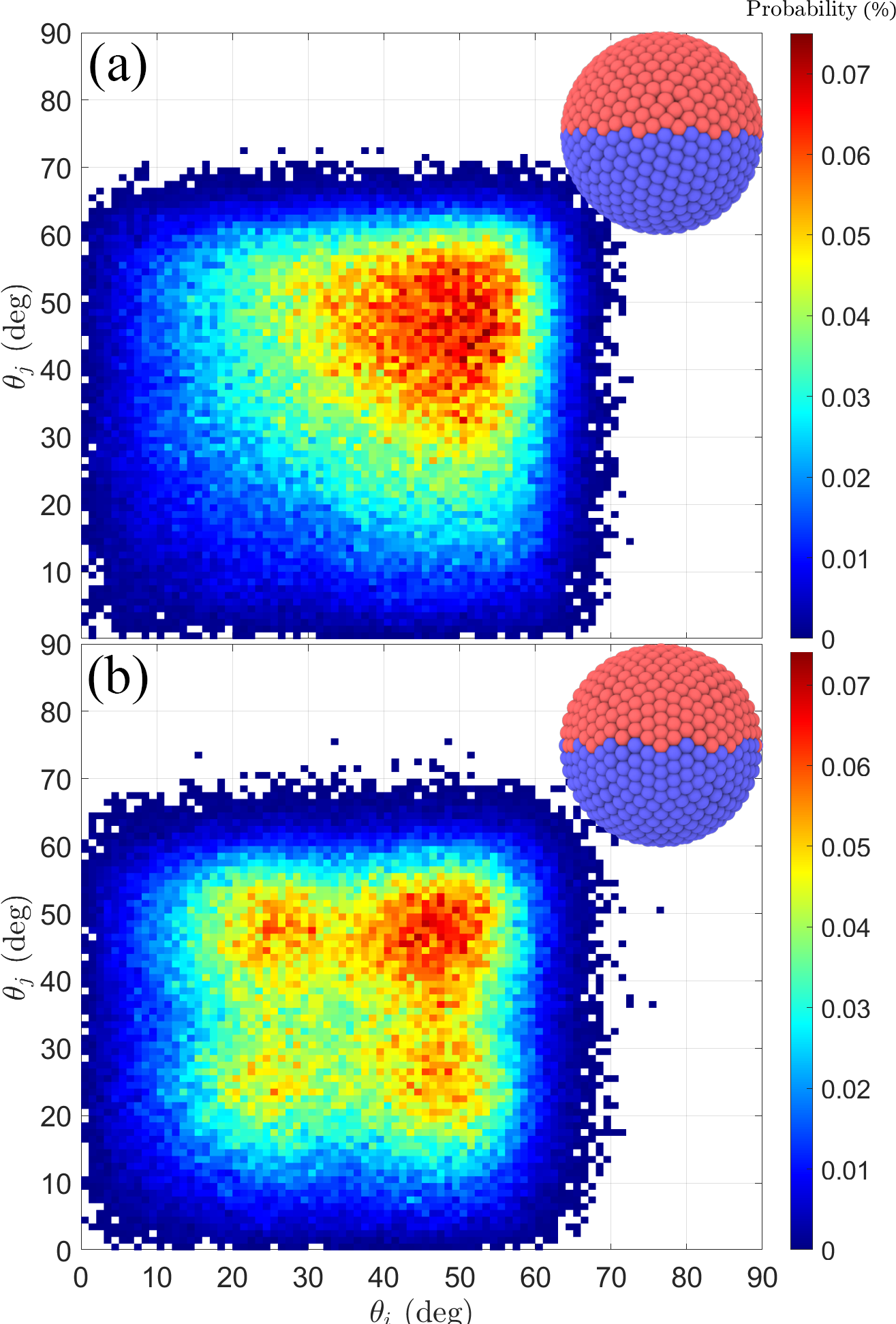}
    \caption{The occurrence probability of $\theta$ angles for spheres on which atoms are distributed (a) randomly, and (b) using a triangulation mesh corresponding to an icosahedron's surface ($T=1.1~\epsilon/k_B$, $A_o=0.05~\epsilon$, $R=10~\sigma$). The designed sphere is shown in each case.}
    \label{Fig_Design_Eff}
\end{figure}

\subsubsection{Sphere design}
\label{sec:4c}

In previous MD research, many-atom JPs have been designed using dodecahedron or icosahedron structures to `evenly' distribute atoms on the surface of the spheres \cite{kharazmi2015diffusion,kobayashi2020structure}. Other workers have homogeneously distributed atoms at different latitudes or longitudes \cite{li2019computer,xu2015self}. However, it is not possible to experimentally synthesise Janus spheres with perfectly ordered structures, nor with a sharp boundary between two hemispheres. Figure \ref{Fig_Design_Eff}(a) shows the polar angle occurrence probability for Janus spheres with atoms distributed randomly on their surfaces, as used in our previous studies \cite{safaei2019molecular,safaei2020stability}. Figure \ref{Fig_Design_Eff}(b) is the occurrence probability for spheres on which the atoms are distributed using a triangulation mesh corresponding to an icosahedron's surface \cite{icosMat}. When the atoms are ordered on the surface, some unexpectedly favoured orientations are predicted. In turn, this ordering could affect the geometries of more complicated nanostructures assembled using these (or similar) spheres.

\subsection{Azimuthal angle}

The occurrence probability distribution for dimer orientations with different azimuthal angles ($\Delta\phi$) is explored in Figs.~\ref{Fig_Phi_Study}(a)-(d). The probability distribution is shown for both NI calculations and MD results at two different temperatures. When using the NI method, an increase in temperature shifts the probability distribution towards higher polar angles (see Sec. \ref{Sec:Temp}) but there is no noticeable effect on the distribution for $\Delta\phi$. However, an unexpected variation in probability was observed when using MD. Smaller values of $\Delta\phi$ were found to be favoured at higher temperatures (Fig. \ref{Fig_Phi_Study}(d)).

This result can be explained by the layering of solvent atoms around the spheres. Less ordering occurs at smaller $\Delta\phi$ angles, increasing the entropy of the system. The free energy for orientations with smaller $\Delta\phi$ is therefore lower, and decreases with increasing temperature.

The layering can be observed in Fig. \ref{Fig_Phi_Study}(e), which shows the spatial distribution of solvent atoms close to hydrophilic surfaces for $\Delta\phi=5\degree$ and $\Delta\phi=175\degree$ when $55\degree < \theta_i, \theta_j < 60\degree$. The first three layers of solvent atoms are approximately the same for both cases, while the fourth and fifth peaks are somewhat stronger at $\Delta\phi=175\degree$. 

Fig. \ref{Fig_Phi_Study}(f) illustrates the reason for this difference. At low $\Delta\phi$, the hydrophilic sides are close to each other so that regions of adjacent layered solvent overlap. Layers are observed at short range $(\lesssim 3~\sigma)$ but the particles each influence longer range ordering of solvent adjacent to their neighbour. At higher $\Delta\phi$, the hydrophilic faces are further away from each other and the layering of solvent atoms is not disturbed by the neighbouring sphere. Near the hydrophobic faces, the range of ordered solvent atoms is much shorter and there is negligible contribution to ordering. 

\subsection{Effect of patchiness}

\begin{figure}[h]
    \includegraphics[width=1\linewidth]{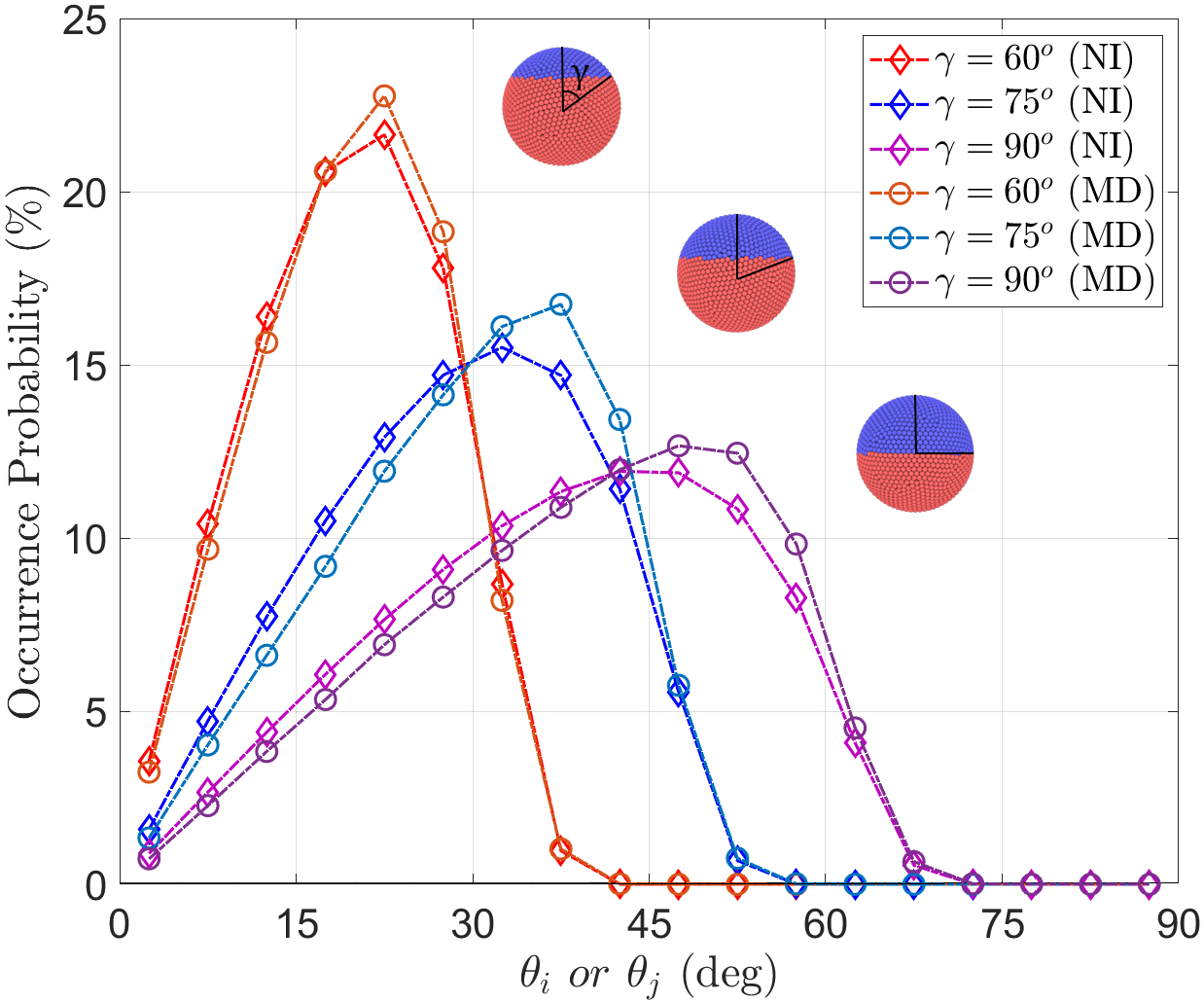}
    \caption{Occurrence probability as a function of polar angle at three different Janus balances ($\gamma$) calculated using both MD and NI methods ($T=1.1~\epsilon/k_B$, $A_o=0.05~\epsilon$, $R=10~\sigma$). Probabilities are averaged over $\Delta\phi$ angles. Insets show the definition of $\gamma$ and Janus spheres with three patchiness values corresponding to the plotted data. Error bars are smaller than the markers.}
    \label{Fig_Patchiness}
\end{figure}

Fig. \ref{Fig_Patchiness} shows the effect of Janus balance on the orientation of Janus spheres. Janus balance ($\gamma$) is defined as the opening angle of the hydrophobic surface (Fig. \ref{Fig_Patchiness}, insets). Both MD and NI results indicate that as $\gamma$ increases, the probability distribution broadens and favours higher polar angles. As the surface area of the hydrophobic side increases, the added configurations that have hydrophobic surfaces shielded from the solvent have high orientational entropy. That is, there is a relatively large number of these configurations. This causes the occurrence probability peak to shift to high angles.

The energies of the configurations added as $\gamma$ increases are similar to those at lower polar angles, as in Fig.~\ref{Fig_Potential_Eff} inset. This is due to the strong short range repulsion of the Lennard-Jones potential. Many configurations do not bring the hydrophilic sides close enough for these interactions to be significant, so most have similar interaction energies.

\section{Conclusion}

To conclude, our calculations using two independent methods have yielded consistent results for the favoured orientations of a dimer of amphiphilic JPs interacting via Lennard-Jones potentials in a solvent. The most favoured configurations are at polar angles between 40 and $55\degree$. Pole-to-pole configurations are unfavoured due to orientational entropy and the short-range nature of the potential, while orientations near $90\degree$ are energetically unfavourable due to solvent exclusion between the particles. Variations in the polar angle probability distribution were explored as a function of the potential, temperature, hydrophobicity, and sphere design.

In comparison with NI results, MD simulations revealed that dynamic changes in the separation between particles and solvent atoms and the nature of the boundary between hemispheres can produce small differences in orientation probabilities. The simulations also revealed that ordering the distribution of atoms on particle surfaces can produce patterns in the probability distribution.

The probability distribution for azimuthal angles ($\Delta\phi$) was uniform when using NI, but this degeneracy was broken due to solvent ordering captured in the MD simulations. The NI method, which directly calculates the interaction energy, is well suited for exploring ranges of practically important parameters, such as different potentials or non-hemispherical patches. Considering the role of orientational entropy, both the NI and MD methods should be useful for designing patches which yield desired `floppy bonds' geometries. The calculation efficiency of these methods suggest that they could be extended to study clusters, and then used to find simplified yet accurate potentials for simulating many-particle structures.

\section{Acknowledgements}

This research was funded by The MacDiarmid Institute for Advanced Materials and Nanotechnology, and simulations were run on the NeSI Mahuika and M\={a}ui Clusters, part of the Centre for eResearch hosted by the University of Auckland. 

\bibliographystyle{apsrev4-1}
\bibliography{Ref.bib}

\end{document}